# Layer-dependent mechanical properties and enhanced plasticity in the van der Waals chromium trihalide magnets

*Fernando Cantos-Prieto[†1], Alexey Falin[†2,3], Martin Alliati[4], Dong Qian[5], Rui Zhang[5], Tao Tao[2], Matthew R. Barnett[3], Elton J. G. Santos\*[6,7], Lu Hua Li\*[3], Efrén Navarro-Moratalla\*[1]*

1. Instituto de Ciencia Molecular, Universitat de València, Calle Catedrático José Beltrán Martínez 2, 46980, Paterna, Spain.

2. Guangdong Provincial Key Laboratory of Functional Soft Condensed Matter, School of Materials and Energy, Guangdong University of Technology, Guangzhou 510006, China.

3. Institute for Frontier Materials, Deakin University, Geelong Waurn Ponds Campus, Waurn Ponds, Victoria 3216, Australia.

4. School of Mathematics and Physics, Queen's University Belfast, BT7 1NN, United Kingdom

5. Department of Mechanical Engineering, The University of Texas at Dallas, Richardson, Texas 75080, USA.

6. Institute for Condensed Matter Physics and Complex Systems, School of Physics and Astronomy, The University of Edinburgh, EH9 3FD, United Kingdom.

7. Higgs Centre for Theoretical Physics, The University of Edinburgh, EH9 3FD, UK

**The mechanical properties of magnetic materials are instrumental for the development of magnetoelastic theories and the optimization of strain-modulated magnetic devices. In particular, two-dimensional (2D) magnets hold promise to enlarge these concepts into the realm of low-dimensional physics and ultrathin devices. However, no experimental study on the intrinsic mechanical properties of the archetypal 2D magnet family of the chromium trihalides has thus far been performed. Here, we report the room temperature layer-dependent mechanical properties of atomically thin $CrCl_3$ and $CrI_3$, finding that the bilayers have Young's moduli of 62.1 GPa and 43.4 GPa, highest sustained strains of 6.49% and 6.09% and breaking strengths of 3.6 GPa and 2.2 GPa, respectively. This portrays the outstanding plasticity of these materials that is qualitatively demonstrated in the bulk crystals. The current study will contribute to the applications of the 2D magnets in magnetostrictive and flexible devices.**

The magnetic moment of a crystal is susceptible to the application of external strain[1], as a consequence magnetostriction has had a big technological relevance in the past century[2–4]. The recent isolation of free-standing 2D magnets[5–9], has settled long-standing fundamental questions[7] and enabled ultrathin magneto-electric devices[10–12]. However, despite recent works that have demonstrated a strong modulation of 2D magnetism in atomically-thin $CrI_3$ under high-pressure values[13,14], direct-strain modulation has only been attempted for < 0.3% strain values[15], and the prospects of the modulation of magnetism in the 2D limit have therefore not been fully explored. This can be attributed to a lack of fundamental understanding of the intrinsic mechanical properties of 2D magnets, which proves vital to realize their various applications. Indeed, although $CrCl_3$ was first studied by Kamerlingh Onnes[16] at the beginning of the last century, no experimental data on the mechanical properties of the magnetic chromium trihalide (CrX$_3$, X = I, Cl, Br) bulk or few-layer crystals has been reported to date.

The mechanical properties of 2D materials have been shown to be different from those of their bulk counterparts. Graphene, for instance, has Young's modulus of ~1 TPa and breaking strength of 130 GPa[17], significantly higher than in graphite[18,19]. A similar trend has been observed in other 2D materials, such as atomically thin hexagonal boron nitride (hBN) (0.87 TPa in Young's modulus and 70 GPa in breaking strength) and molybdenum disulfide (MoS$_2$) (0.33 TPa in Young's modulus and 30 GPa in breaking strength)[20,21]. It is worth mentioning that these strength values are far beyond the yield strength measured in conventional materials (i.e. ~3 GPa for that of silicon)[22], demonstrating the capability of 2D materials to sustain an enormous strain without failure[23], e.g. up to 25% in the case of graphene[17]. On the other hand, the multilayer forms of these van der Waals crystals can benefit from their layered structure to achieve large plasticity. Such exceptional behavior has been recently reported in InSe, portraying this material as a strong candidate for near-future deformable electronics[24,25]. It is therefore timely to explore the layer-dependent intrinsic mechanical properties of the chromium trihalides as archetypal magnetic 2D materials.

In our experiment, we obtained atomically-thin CrI$_3$ and CrCl$_3$ flakes down to the bilayer (2L) by mechanical exfoliation of bulk crystals. The exfoliation was directly performed on substrates with pre-fabricated microwells for atomic force microscopy (AFM) nanoindentation (see Supporting Information for details). Figure 1a,d shows an optical micrograph of atomically-thin CrI$_3$ and CrCl$_3$ covering several holes in a 90 nm-SiO$_2$/Si substrate. Figure 1b,e shows the corresponding AFM images in contact mode, portraying a thickness of 1.7 nm and 1.4 nm (Figure 1c,f), respectively, which corresponds to 2L CrCl$_3$ and CrI$_3$.

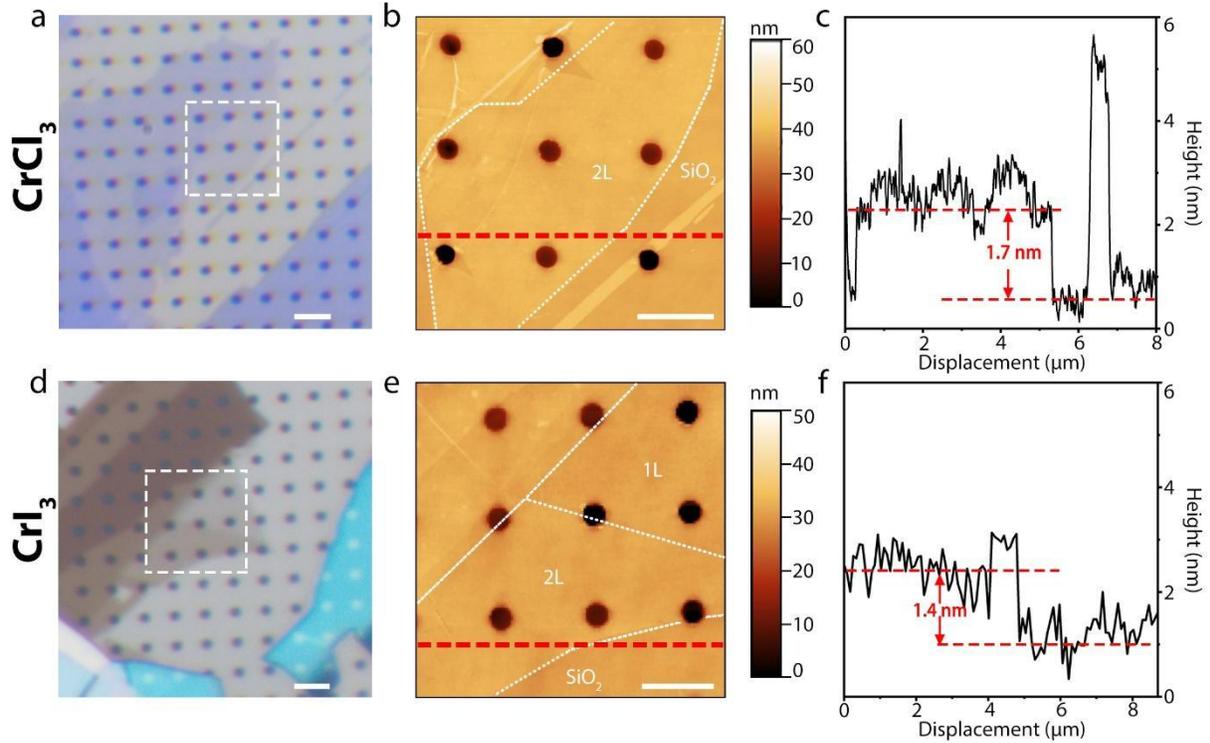

**Figure 1. Characterization of CrX$_3$ nanosheets. a, d,** Optical microscopy image of 2L and few-layers of CrCl$_3$ and CrI$_3$ crystals, respectively, on a SiO2/Si substrate suspended over micro-wells of 600 nm in diameter; **b, e,** AFM image of the CrCl$_3$ and CrI$_3$ thin crystals, respectively, corresponded to the square area of optical images (a,d, respectively); **c, f,** the corresponding height traces of the dashed line in **b,e** of 2L CrCl$_3$ and 2L CrI$_3$ crystals, respectively. Scale bars in white, 3 μm in **a** and **d**, 2 μm in **b** and **e**.

The mechanical properties of the few-layer CrI$_3$ and CrCl$_3$ were probed by the nanoindentation technique performed with the same AFM used for topographic inspection[17,20]. The load-displacement curves were obtained by applying a load at the center of each suspended region until fracture for a minimum of five indentations per thickness per material to ensure the repeatability of the results. The curves were then fitted by a well-established model[17] (see section 4 of the Supporting Information) as demonstrated in Figure 2a. From these results, we extracted Young's modulus (E) for both materials in terms of their layer count (Figure 2b). The breaking strength (σ) (Figure 2c) and ultimate strain values were determined based on the obtained fracture loads and the load-displacement relationships by means of finite element simulation (FEM) (see section 5 in the Supporting Information). The volumetric Young's modulus and breaking strength of 2L $CrI_3$ and 2L $CrCl_3$ were E = 43.4 ± 4.4 GPa and σ = 2.2 ± 0.5 GPa and E = 62.1 ± 4.8 GPa and σ = 3.6 ± 0.4 GPa, respectively. The ultimate strain was found directly under the tip and its values for 2L $CrI_3$ and 2L $CrCl_3$ were 6.09% and 6.49%, respectively. A direct comparison between the two materials indicates that both the Young's modulus and the breaking strength of $CrCl_3$ were larger than those of $CrI_3$, depicting that the chromium trihalide materials with a heavier halide exhibit a lower mechanical stiffness. This trend correlates nicely with the ionic character of the Cr-X bond, which is stronger in the Cr-Cl interaction compared to Cr-I [26].

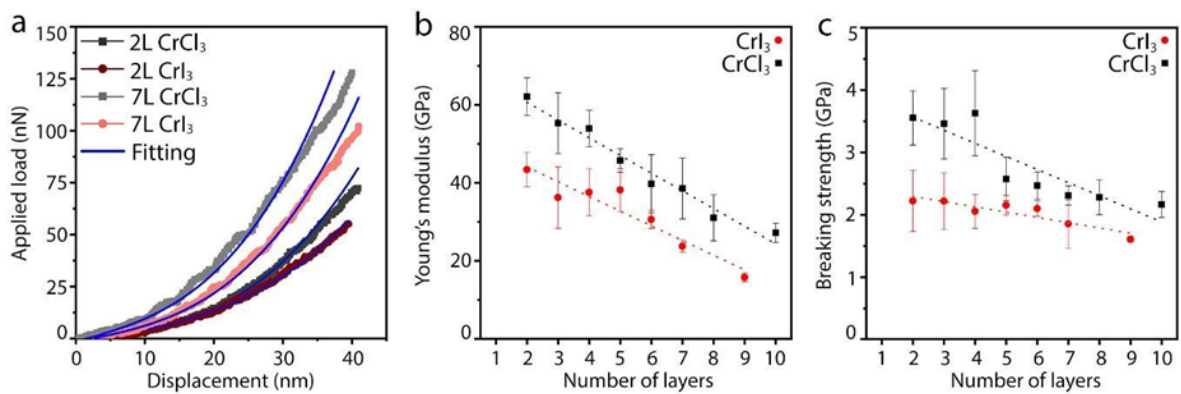

**Figure 2. Mechanical properties of $CrX_3$ nanosheets. a**, Load–displacement curves and the corresponding fittings for 2L,7L for $CrCl_3$ and $CrI_3$; **b**, Volumetric Young's modulus and **c**, breaking strength of $CrCl_3$ and $CrI_3$ crystals of different thicknesses, along with dashed lines that show the linear fit of experimental Young's moduli values.

Remarkably, as the thickness of the flakes increases, both atomically-thin materials show a drop in Young's modulus and breaking strength. For example, 9L CrI$_3$ had E = 15.8 ± 1.2 GPa and σ =1.6 ± 0.04 GPa, representing a 64% and 27% decrease in Young's modulus and breaking strength compared to 2L, respectively. Similar trends were observed in CrCl$_3$, where 10L CrCl$_3$ had E = 27.1±2.5 GPa and σ = 2.2± 0.2GPa. In order to provide further insights into the layer-dependent mechanical properties of the chromium trihalides, we have undertaken van der Waals-corrected density functional theory (vdW-DFT) calculations to unveil the energy landscape of interlayer sliding shifts during the mechanical tests (see section 6 in the Supporting Information for details). Figure 3a shows a schematic of the atomic structure of bilayer CrX$_3$ along the crystallographic c axis (direction of the AFM tip motion during indentation) in the monoclinic phase (space group C2/m) present at room temperature, with the definition of the two interlayer sliding paths utilized in the simulations: along [100] and along [010]. We apply a fractional lateral shift on one chromium-trihalide layer relative to the other starting from the AB stacking order (Figure 3b,c). The FEM simulations predict that the suspended 2L CrX$_3$ crystals are mostly under small in-plane strain in the area far from the contact region even under the fracture loads. This picture changes in the region close to the indentation center where out-of-plane compression starts to play a key role in the fracture mechanism. Figures 3d, 3g show the in-plane strain (solid lines) and out-of-plane compression (dashed lines) distributions close to the indentation center under different fracture loads for 2L CrCl$_3$ and 2L CrI$_3$, respectively. On the vdW-DFT calculations, three distinct regions were chosen to evaluate the sliding energy barriers. In the region far away from the indentation center, the equilibrium interlayer interaction occurs at 0 GPa out-of-plane compression and 0% in-plane strain (i.e. 0 GPA and 0% for both CrCl$_3$ and CrI$_3$). This choice of strain conditions is a valid approximation to our experiments, where extremely low values of strain, < 0.5%, are found at the membrane edges (see Supporting Figure S3). The area just outside of the contact region is under a large in-plane strain but without any out-of-plane compression (5.35% and 0 GPa for $CrCl_3$; 4.79% and 0 GPa for $CrI_3$). The tip contact region experiences the highest in-plane strain and out-of-plane compression under the fracture loads (0.49 GPa and 6.49% for $CrCl_3$; 0.36 GPa and 6.09% for $CrI_3$). Figures 3e, 3f, 3h, 3i summarize the sliding energy per formula unit obtained for $CrI_3$ and $CrCl_3$ at different values of interlayer pressure and in-plane strain as provided by FEM simulations. In the regions of the membranes beyond ~8.5 nm from the indentation centers (see Table S2), where no pressure and small strain are present in the systems, the individual layers of 2L $CrI_3$ and $CrCl_3$ tend to slide over each other despite the path considered, i.e. [100] or [010] (Figure 3e-f, 3h-i). This process is mediated by thermal fluctuations ($kT$ = 25.7 meV) which are present at room temperature. The interlayer barriers are below $kT$ for the majority of the positions with the only exception at the fractional shift of 2/3. At this crystallographic position, there is a slight increment of the energy above $kT$, which prevents further sliding along both [100] and [010]. This indicates that the layers can displace almost freely with little energetic opposition (Figure 3f,i). As pressure and strain are applied (see Table S2), there is an increment of the energetic barriers at 2/3 along [100] for $CrCl_3$ (168 meV) and $CrI_3$ (209 meV) which indicates that the layers may find difficulties to slide over at that particular position (Figure 3e,h). The main driving force for such enhancement of the energies is the strong overlap of the charge density at 2/3 (Figure S5). Conversely, along [010] at 0.49 GPa and 6.49 %, and 0.36 GPa and 6.09% for both $CrCl_3$ and $CrI_3$, respectively, the energies at 2/3 and their multiple positions (0, 1/3, 1) are below kT (Figure 3f, 3i) but slightly increments at intermediate positions (1/6, 1/2, 5/6) although still smaller than the barrier at 2/3 along [100] at finite strain and pressure. This suggests that the layers may choose a combination of sliding paths to minimize their energies as the pressure is applied. That is, the layers may start along the path [100] but may change direction to [010] to minimize their energies. Since the sliding path from 1/2 to 2/3 along [100] is symmetrically the same as along [010] (see Supporting Figure S7), the layers will follow a downhill energy profile from ~76 meV ($CrCl_3$) and ~98 meV ($CrI_3$) to 0 meV on both cases rather than increase their energy following the same path along [100]. These results are consistent with the variation of mechanical properties versus the number of layers which follows our previous analysis on graphene and hBN[20] providing a plausible explanation for the layer-dependence of the mechanical properties in $CrX_3$. It is worth mentioning that the energetic barriers observed in the sliding of the layers are particularly sensitive to the relaxation of the atoms involved (Cr, Cl, I) during the computation. Supporting Figure S8 shows results without any relaxation in the layers, which resulted in larger barriers. Indeed, lower energies than those shown in Figure 3e-f and Figure 3h-i can be achieved when relaxation of the Cr atoms are also taken into account (see Supporting Figure S9) with a consequent expansion of the interlayer distance (see Supporting Figure S10). This correlates well with the positions where the energies increase during the sliding and suggests that changes of stacking order should be followed by expansion or contraction of the interlayer distance as recently measured[14]. In addition, the magnitudes of the barriers for $CrI_3$ are moderately larger than those for $CrCl_3$ which suggest a slightly more stable dependence of the mechanical properties with the thickness, in agreement with the overall experimental trend observed.

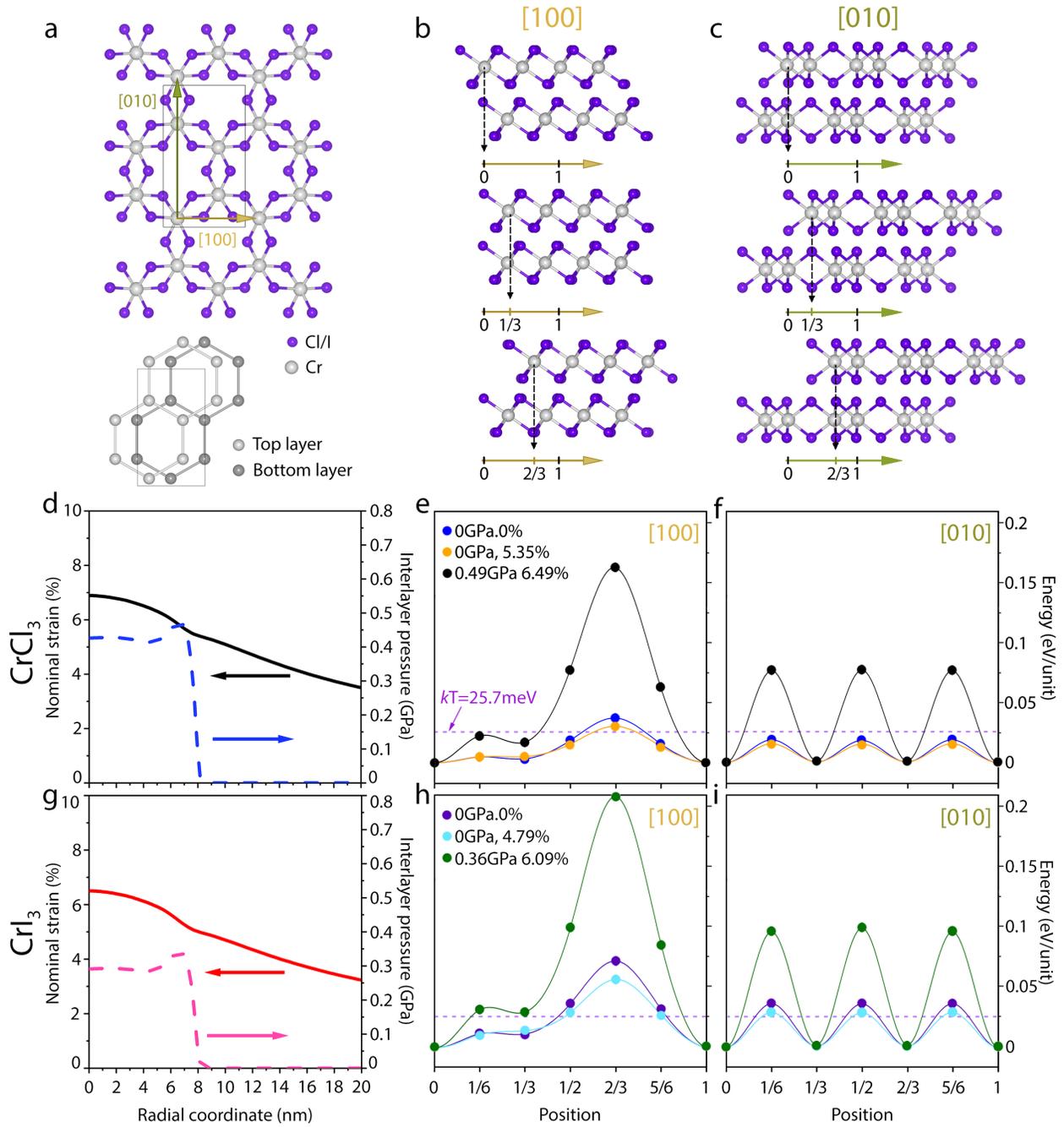

**Figure 3. Sliding energies of bilayer CrX3 under different in-plane strain and out-of-plane compression conditions. a,** Top-view of the bilayer structure utilized in the vdW-DFT simulations, respectively. Two high-symmetry directions along the [100] and [010] are considered as representatives of the lateral sliding process occurring in the structures. Only Cr atoms are

shown in a bottom to highlight the bilayer stacking. **b-c**, The different positions correspond to fractional lateral shifts of the top-layer relative to the original AB stacking in units of [1/6, 0] and [1/6, -1/6] over the unit cell along [100] and [010], respectively. **d,** FEM calculations of nominal strain (solid line) and interlayer pressure (dashed line) in 2L $CrCl_3$ within a radial distance of 20 nm from the indentation center, where three distinct regions were chosen to study sliding energy per formula unit (eV/unit) of bilayer $CrCl_3$ along e, [100] and f, [010], respectively. The sliding energy was determined with the Cr atoms being fixed and the Cl atoms being relaxed in the simulation (see Supporting Figures S8-S10 for additional details). The dashed line indicates available thermal-energy at room temperature ($kT$ =25.7 meV). Colored dots are the calculated vdW-DFT energies with a cubic interpolation (solid lines) between different positions. **g-i**, Analogous analysis as in d-f for 2L $CrI_3$. The monoclinic (space group C2/m) stacking order was utilized in all simulations.

Overall, the measured mechanical values are among the smallest ones observed within the family of 2D materials, ie. much less stiff than 2D transition metal dichalcogenides and mica[21,27,28]. Figure 4 shows a map of the mechanical properties of atomically-thin $CrX_3$ compared to other materials. The position of $CrX_3$ on this chart can be qualitatively explained by taking into account the bonding energies inside of the crystal, which scale according to the magnitudes of Young's modulus and breaking strength. While the dissociation energy for the honeycomb of C atoms in graphene yields a value of 805 kJ/mol[29], our DFT calculations indicate a formation energy of 260.9 kJ/mol for $CrI_3$ and 597.7 kJ/mol for $CrCl_3$, depicting a weaker interatomic interaction than that of the graphene lattice. In addition, within the chromium halide family, the smaller the ionic character the larger the bond energies[26], with a variation of the electronic localization function across the different Cr-halides[30]. These results underline the soft nature of the chromium trihalides, which makes them extremely sensitive to small stress changes and very effective for strain modulation.

Although these results place the chromium trihalides as one of the softest 2D materials that have been experimentally measured so far, their breaking strength up to 10L is larger than that of silicon (~2 GPa)[22], showcasing the general outstanding mechanical properties of 2D materials. It is also important to consider the significance of the presence of imperfections in the crystals, which affects the elastic behavior. Griffith described how the breaking strength in brittle materials (see also section 7 on the Supporting information) is governed by defects and imperfections[31], establishing a limit of $\sigma \sim E/9$ by experimental extrapolation. In the limit of an ideal material mechanics are governed by its molecular tensile strengths. In both chromium trihalides, the bilayers ($\sigma \sim E/20$) and the multi-layers ($\sigma \sim E/10$) follow a behavior close to this limit. These results suggest that the mechanical behavior in $CrX_3$ thin crystals is determined by the interatomic interactions rather than defects, indicating a high crystallinity and a low density of impurities in the suspended regions. In comparison, polycrystalline classical materials like silicon[22] or tungsten alloys[32], $\sigma < E/100$, report much lower values[31]. The nonlinear elastic constitutive behavior was assumed for modeling $CrX_3$ few layer crystals in FEM, and the derived maximum strains are close to ~6-6.5% for the bilayers (see section 5 on the Supporting information). The prospects of the combination of the exceptional flexibility and strengths with the intrinsic magnetism of atomically-thin $CrX_3$ nature hold promise for an enhanced strain-tunability in ultrathin magneto-mechanical devices[33].

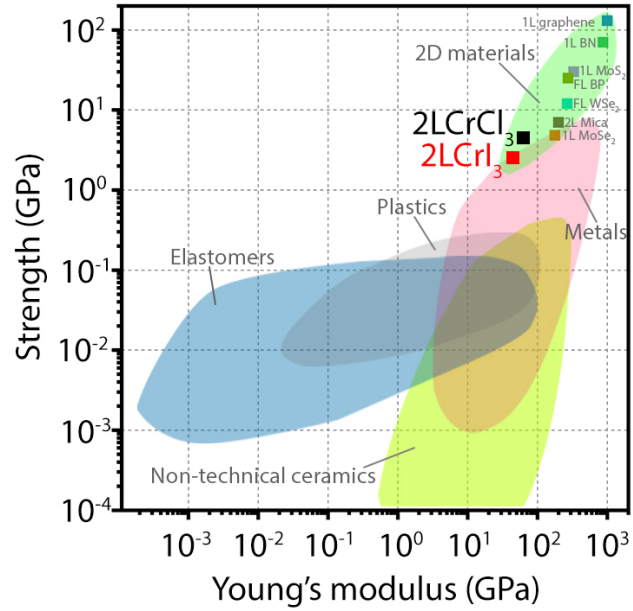

**Figure 4. Map of mechanical properties of different materials in Young's modulus - Ultimate/breaking strength space.** The mechanical properties of different types of materials, including 2D crystals and measured here 2L of $CrX_3$, are compared.

Considering the remarkable flexibility of few-layer $CrCl_3$ and $CrI_3$, and the interlayer sliding origin of the layer-dependent Young's moduli, we investigated the plastic behavior of the two magnetic van der Waals materials in their bulk form. This property is of great relevance for future flexible devices, and it has recently been observed in bulk crystals of InSe[25]. The deformability factor ($\Xi$) proposed by Wei T-R et al. can be useful as a way to frame the plastic behavior of a material, it is related to the sliding ($E_s$) and cleavage ($E_c$) energies of layered materials via:

$$\Xi = (E_c/E_s)\cdot(1/E) \tag{1}$$

where E is the volumetric Young's modulus. The magnitudes of $\Xi$ for bulk $CrCl_3$ and $CrI_3$ are plotted in Figure 5a,b as a function of the Young modulus and bandgaps for different materials with different electronic properties (semiconducting, insulators and metals). The cleavage energies were defined as the energies to separate bilayer $CrX_3$ systems to two monolayers (see Supporting Figure S6), the sliding energies were taken from the most energetically favored sliding path in the equilibrium state, i.e. along [010] direction (Figure 3), and the Young's modulus values were extracted from the experimental data for 2L $CrCl_3$ and $CrI_3$. For both bulk $CrCl_3$ and $CrI_3$ the cleavage energies are larger than their sliding energies (see Table S3). Interestingly, the magnetic $CrX_3$ showed one of the highest deformability factors of the 2D materials, even larger than that of the recently reported InSe (Figure 5)[25]. This outstanding capability for deformation is experimentally illustrated by macroscopically folding bulk $CrI_3$ and $CrCl_3$ crystals in Figure 5c. Upon further testing, $CrX_3$ multilayered crystals could be confirmed to exhibit a superplastic behavior, which would open the door for their use in easily deformable and flexible devices that incorporate intrinsic magnetism.

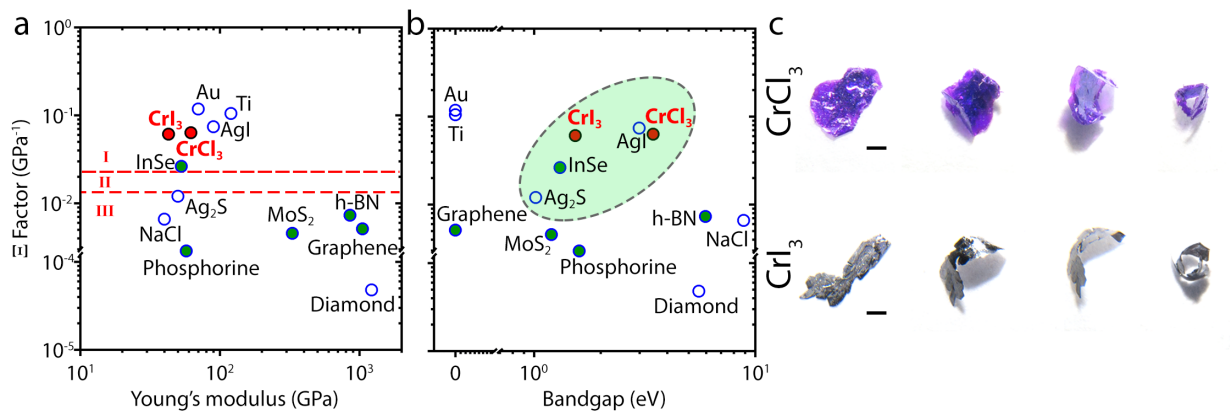

**Figure 5. Deformability factor and enhanced plastic behavior of multilayer CrX3. a,** Deformability factor dependence on the Young Modulus, Regions I, II and III correspond to Plastic-flexible, Potentially deformable and Brittle-rigid regions, respectively. The layered van der Waals materials are shown as green symbols, in red are our experimental results. **b,** Deformability

factor dependence on the bandgap for the same materials as panel a. The dashed line encircled green area are materials that show exceptional plastic behavior. The bandgaps for $CrCl_3$ and $CrI_3$ nanosheets are 3.44 and 1.53 eV, respectively[33]. Panel c shows a folding sequence of flat bulk crystals of $CrCl_3$ (top strip) and $CrI_3$ (bottom strip) into a ring-like structure (enlarged in Figure S12). The scale bars are 1 mm.

**Correspondence to:**

Elton J. Santos - Higgs Centre for Theoretical Physics, The University of Edinburgh, EH9 3FD, UK; Email: esantos@ed.ac.uk

Lu Hua Li - Institute for Frontier Materials, Deakin University, Geelong Waurn Ponds Campus, Waurn Ponds, Victoria 3216, Australia; Email: luhua.li@deakin.edu.au

Efrén Navarro-Moratalla - Instituto de Ciencia Molecular, Universitat de València, Calle Catedrático José Beltrán Martínez 2, 46980, Paterna, Spain; Email: efren.navarro@uv.es

**Author Contributions**

The manuscript was written through contributions of all authors. All authors have given approval to the final version of the manuscript. †These authors contributed equally.

**Acknowledgments**

EJGS acknowledges computational resources through the UK Materials and Molecular Modelling Hub for access to THOMAS supercluster, which is partially funded by EPSRC (EP/P020194/1); CIRRUS Tier-2 HPC Service (ec131 Cirrus Project) at EPCC (http://www.cirrus.ac.uk) funded by the University of Edinburgh and EPSRC (EP/P020267/1); ARCHER UK National Supercomputing Service (http://www.archer.ac.uk) via Project d429. EJGS acknowledges the EPSRC Early Career Fellowship (EP/T021578/1) and the University of Edinburgh for funding support. ENM acknowledges the European Research Council (ERC) under the Horizon 2020 research and innovation programme (ERC StG, grant agreement No. 803092) and to the Spanish Ministerio de Ciencia, Innovación y Universidades for financial support from the Ramon y Cajal program (Grant No. RYC2018-024736-I). FCP acknowledges the same institution for the FPU program (Grant No. FPU17/01587). This work was also supported by the Spanish Unidad de Excelencia "María de Maeztu" (CEX2019-000919-M). FCP and ENM are particularly grateful to the microscopy service at the SCSIE (Universitat de Valencia) and the Electron Microscopy Service at the Polytechnic University of Valencia for their technical support.

**References**

(1) Brown, W. F. Theory of Magnetoelastic Effects in Ferromagnetism. J. Appl. Phys. **1965**, 36 (3), 994–1000.

(2) Meydan, T.; Oduncu, H. Enhancement of Magnetorestrictive Properties of Amorphous Ribbons for a Biomedical Application. Sens. Actuators A Phys. **1997**, 59 (1), 192–196.

(3) Bieńkowski, A.; Szewczyk, R. The Possibility of Utilizing the High Permeability Magnetic Materials in Construction of Magnetoelastic Stress and Force Sensors. Sens. Actuators A Phys. **2004**, 113 (3), 270–276.

(4) Bieńkowski, A.; Szewczyk, R.; Salach, J. Industrial Application of Magnetoelastic Force and Torque Sensors. Acta Phys. Pol. A **2010**, 118 (5), 1008–1009.

(5) Gong, C.; Li, L.; Li, Z.; Ji, H.; Stern, A.; Xia, Y.; Cao, T.; Bao, W.; Wang, C.; Wang, Y.; Qiu, Z. Q.; Cava, R. J.; Louie, S. G.; Xia, J.; Zhang, X. Discovery of Intrinsic Ferromagnetism in Two-Dimensional van Der Waals Crystals. Nature **2017**, 546 (7657), 265–269.

(6) Huang, B.; Clark, G.; Navarro-Moratalla, E.; Klein, D. R.; Cheng, R.; Seyler, K. L.; Zhong, D.; Schmidgall, E.; McGuire, M. A.; Cobden, D. H.; Yao, W.; Xiao, D.; Jarillo-Herrero, P.; Xu, X. Layer-Dependent Ferromagnetism in a van Der Waals Crystal down to the Monolayer Limit. Nature **2017**, 546 (7657), 270–273.

(7) Lee, J.-U.; Lee, S.; Ryoo, J. H.; Kang, S.; Kim, T. Y.; Kim, P.; Park, C.-H.; Park, J.-G.; Cheong, H. Ising-Type Magnetic Ordering in Atomically Thin FePS 3. Nano Lett. **2016**, 16, 12, 7433–7438 .

(8) Wahab, D. A.; Augustin, M.; Valero, S. M.; Kuang, W.; Jenkins, S.; Coronado, E.; Grigorieva, I. V.; Vera-Marun, I. J.; Navarro-Moratalla, E.; Evans, R. F. L.; Novoselov, K. S.; Santos, E. J. G. Quantum Rescaling, Domain Metastability, and Hybrid Domain-Walls in 2D CrI3 Magnets. Adv. Mater. **2021**, 33 (5), e2004138.

(9) Augustin, M.; Jenkins, S.; Evans, R. F. L.; Novoselov, K. S.; Santos, E. J. G. Properties and Dynamics of Meron Topological Spin Textures in the Two-Dimensional Magnet CrCl3. Nat. Commun. **2021**, 12 (1), 185.

(10) Gong, C.; Zhang, X. Two-Dimensional Magnetic Crystals and Emergent Heterostructure Devices. Science **2019**, 363 (6428).

(11) Huang, B.; Clark, G.; Klein, D. R.; MacNeill, D.; Navarro-Moratalla, E.; Seyler, K. L.; Wilson, N.; McGuire, M. A.; Cobden, D. H.; Xiao, D.; Others. Electrical Control of 2D Magnetism in Bilayer CrI 3. Nat. Nanotechnol. **2018**, 13 (7), 544–548.

(12) Jiang, S.; Li, L.; Wang, Z.; Mak, K. F.; Shan, J. Controlling Magnetism in 2D CrI3 by Electrostatic Doping. Nat. Nanotechnol. **2018**, 13 (7), 549–553.

(13) Li, T.; Jiang, S.; Sivadas, N.; Wang, Z.; Xu, Y.; Weber, D.; Goldberger, J. E.; Watanabe, K.; Taniguchi, T.; Fennie, C. J.; Others. Pressure-Controlled Interlayer Magnetism in Atomically Thin CrI 3. Nat. Mater. **2019**, 18 (12), 1303–1308.

(14) Song, T.; Fei, Z.; Yankowitz, M.; Lin, Z.; Jiang, Q.; Hwangbo, K.; Zhang, Q.; Sun, B.; Taniguchi, T.; Watanabe, K.; McGuire, M. A.; Graf, D.; Cao, T.; Chu, J.-H.; Cobden, D. H.; Dean, C. R.; Xiao, D.; Xu, X. Switching 2D Magnetic States via Pressure Tuning of Layer Stacking. Nat. Mater. **2019**, 18 (12), 1298–1302.

(15) Jiang, S.; Xie, H.; Shan, J.; Mak, K. F. Exchange Magnetostriction in Two-Dimensional Antiferromagnets. Nat. Mater. **2020**, 19, 1295–1299.

(16) Woltjer, H. R.; Kamerlingh Onnes, H. Further Experiments with Liquid Helium. Z. Magnetic Researches. XXVIII. Magnetisation of Anhydrous CrC/ COC/2 and NiC/ at Very /ow Temperatures. Leiden Communications **1925**, 173b.

(17) Lee, C.; Wei, X.; Kysar, J. W.; Hone, J. Measurement of the Elastic Properties and Intrinsic Strength of Monolayer Graphene. Science **2008**, 321 (5887), 385–388.

(18) Cost, J. R.; Janowski, K. R.; Rossi, R. C. Elastic Properties of Isotropic Graphite. The Philosophical Magazine: A Journal of Theoretical Experimental and Applied Physics **1968**, 17 (148), 851–854.

(19) Blakslee, O. L.; Proctor, D. G.; Seldin, E. J.; Spence, G. B.; Weng, T. Elastic Constants of Compression-Annealed Pyrolytic Graphite. J. Appl. Phys. **1970**, 41 (8), 3373–3382.

(20) Falin, A.; Cai, Q.; Santos, E. J. G.; Scullion, D.; Qian, D.; Zhang, R.; Yang, Z.; Huang, S.; Watanabe, K.; Taniguchi, T.; Barnett, M. R.; Chen, Y.; Ruoff, R. S.; Li, L. H. Mechanical Properties of Atomically Thin Boron Nitride and the Role of Interlayer Interactions. Nat. Commun. **2017**, 8, 15815.

(21) Bertolazzi, S.; Brivio, J.; Kis, A. Stretching and Breaking of Ultrathin MoS2. ACS Nano **2011**, 5 (12), 9703–9709.

(22) Tsuchiya, T. Tensile Testing of Silicon Thin Films. Fatigue Fract. Eng. Mater. Struct. **2005**, 28 (8), 665–674.

(23) Castellanos-Gomez, A.; van Leeuwen, R.; Buscema, M.; van der Zant, H. S. J.; Steele, G. A.; Venstra, W. J. Single-Layer MoS(2) Mechanical Resonators. Adv. Mater. **2013**, 25 (46), 6719–6723.

(24) Zhao, Q.; Frisenda, R.; Wang, T.; Castellanos-Gomez, A. InSe: A Two-Dimensional Semiconductor with Superior Flexibility. Nanoscale **2019**, 11 (20), 9845–9850.

(25) Wei, T.-R.; Jin, M.; Wang, Y.; Chen, H.; Gao, Z.; Zhao, K.; Qiu, P.; Shan, Z.; Jiang, J.; Li, R.; Chen, L.; He, J.; Shi, X. Exceptional Plasticity in the Bulk Single-Crystalline van Der Waals Semiconductor InSe. Science **2020**, 369 (6503), 542–545.

(26) Dean, J. A. Lange's Handbook of Chemistry; New york; London: McGraw-Hill, Inc., 1999.

(27) Castellanos-Gomez, A.; Poot, M.; Amor-Amorós, A.; Steele, G. A.; van der Zant, H. S. J.; Agraït, N.; Rubio-Bollinger, G. Mechanical Properties of Freely Suspended Atomically Thin Dielectric Layers of Mica. Nano Res. **2012**, 5 (8), 550–557.

(28) Liu, K.; Yan, Q.; Chen, M.; Fan, W.; Sun, Y.; Suh, J.; Fu, D.; Lee, S.; Zhou, J.; Tongay, S.; Ji, J.; Neaton, J. B.; Wu, J. Elastic Properties of Chemical-Vapor-Deposited Monolayer MoS2, WS2, and Their Bilayer Heterostructures. Nano Lett. **2014**, 14 (9), 5097–5103.

(29) Costescu, B. I.; Baldus, I. B.; Gräter, F. Graphene Mechanics: I. Efficient First Principles Based Morse Potential. Phys. Chem. Chem. Phys. **2014**, 16 (24), 12591–12598.

(30) Kartsev, A.; Augustin, M.; Evans, R. F. L.; Novoselov, K. S.; Santos, E. J. G. Biquadratic Exchange Interactions in Two-Dimensional Magnets. npj Computational Materials **2020**, 6 (1), 150.

(31)     Griffith, A. A.; Taylor, G. I. VI. The Phenomena of Rupture and Flow in Solids. Philosophical Transactions of the Royal Society of London. Series A, Containing Papers of a Mathematical or Physical Character **1921**, 221 (582-593), 163–198.

(32)     Gong, X.; Fan, J.; Ding, F. Tensile Mechanical Properties and Fracture Behavior of Tungsten Heavy Alloys at 25--1100° C. Materials Science and Engineering: A **2015**, 646, 315–321.

(33)     Zhang, W.-B.; Qu, Q.; Zhu, P.; Lam, C.-H. Robust Intrinsic Ferromagnetism and Half Semiconductivity in Stable Two-Dimensional Single-Layer Chromium Trihalides. J. Mater. Chem. **2015**, 3 (48), 12457–12468.

(34)     Michael A. McGuire, Hemant Dixit, Valentino R. Cooper, and Brian C. Sales. Coupling of Crystal Structure and Magnetism in the Layered, Ferromagnetic Insulator CrI3. Chemistry of Materials **2015** 27 (2), 612-620.

(35)     McGuire, M. A.; Clark, G.; Santosh, K. C.; Chance, W. M.; Jellison, G. E., Jr; Cooper, V. R.; Xu, X.; Sales, B. C. Magnetic Behavior and Spin-Lattice Coupling in Cleavable van Der Waals Layered $CrCl_3$ Crystals. Physical Review Materials **2017**, 1 (1), 014001.

(36)     Tkatchenko, A.; Scheffler, M. Accurate Molecular van Der Waals Interactions from Ground-State Electron Density and Free-Atom Reference Data. Phys. Rev. Lett. **2009**, 102 (7), 073005.

(37)     Dudarev, S. L.; Botton, G. A.; Savrasov, S. Y.; Humphreys, C. J.; Sutton, A. P. Electron-Energy-Loss Spectra and the Structural Stability of Nickel Oxide: An LSDA+U Study. Phys. Rev. B Condens. Matter **1998**, 57 (3), 1505–1509.

(38)     Liu, J.; Sun, Q.; Kawazoe, Y.; Jena, P. Exfoliating Biocompatible Ferromagnetic Cr-Trihalide Monolayers. Phys. Chem. Chem. Phys. **2016**, 18 (13), 8777–8784.

(39)     Cadelano, E.; Palla, P. L.; Giordano, S.; Colombo, L. Elastic Properties of Hydrogenated Graphene. Phys. Rev. B Condens. Matter **2010**, 82 (23), 235414.

# Supporting information

# Contents

1. **Crystal growth**

The single crystals of CrI$_3$ were grown by chemical vapor transport. Chromium powder (99.996, Alfa-Aesar) and beads of iodine (anhydrous 99.999 %, Sigma-Aldrich) were mixed in a 1:3 ratio inside an argon atmosphere in a glovebox. 972 mg of the mixture were loaded into a silica ampoule with a length, inner diameter and outer diameter of 500 mm, 15 mm and 16 mm respectively. The ampoule was extracted from the glovebox with a ball valve covering the open end to prevent air exposure and then it was immediately evacuated using a turbomolecular pump down to $6 \cdot 10^{-6}$. Once the pressure stabilized, the close end was dipped in liquid nitrogen to prevent the sublimation of the iodide beads. The ampoule was then flame sealed and introduced into a three-zone furnace with the material in the leftmost zone. The other two zones were heated up from room temperature to 650°C in 1440 minutes and kept for 1620 minutes to minimize nucleation sites in the growth zone. Then, the leftmost side was heated up to 700°C in 180 minutes. The three-zone furnace had a temperature gradient of 700°C/650°C /675°C. Subsequently, the temperature was kept constant for 7 days and cooled down naturally. Phase purity was inspected by powder X-ray diffraction (Fig. S1). After the crystal growth, the quartz tube was transferred into an argon glove box to prevent hydration, where the crystals were first purified by sublimation to remove the halide excess. Crystals were ground into powder using a razor blade and inserted into a capillary to perform X-Ray diffraction analysis. The phase of the bulk crystals was confirmed by fitting the X-ray diffraction pattern to the patterns computed from the single crystal structures previously reported[34] (Fig. S1). The fit of the X-ray pattern revealed a monoclinic crystal system with a C12/m1 space group consistent with the crystal structure previously reported for CrI$_3$ at room temperature (ICSD 251654). Crystals of CrCl$_3$ were commercially obtained from © Strem chemicals as Chromium(III) chloride, anhydrous (99.9%-Cr), CAS number: 10025-73-7. A representative CrI$_3$ grown crystal is shown in Figure S1 together with its XRD characterization and a commercial CrCl$_3$ crystal.

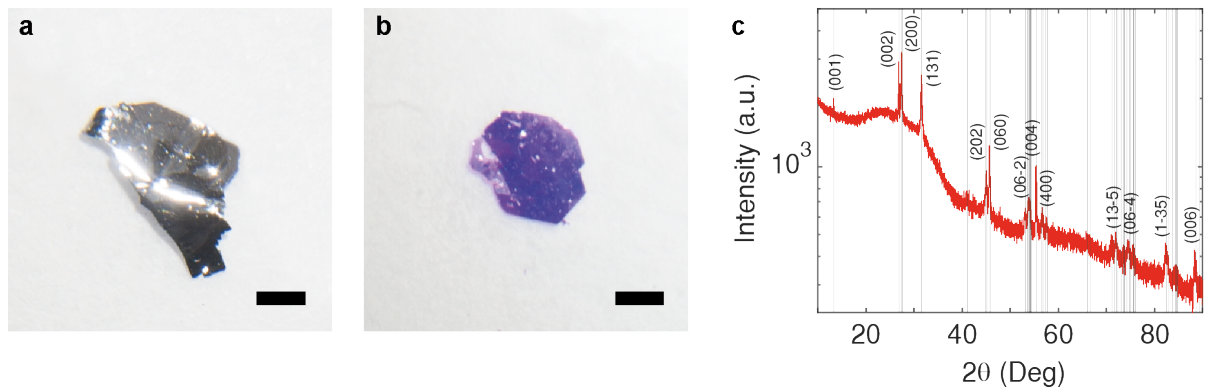

**Figure S1. CrI3 (a) and CrCl3 (b) starting bulk crystals.** Panel **c** shows the XRD powder pattern of CrI$_3$. The continuous red line corresponds to the measured diffraction points. Vertical lines indicate the theoretical positions of the Bragg diffraction peaks computed from the ICSD 251654 database, the miller indices of the main peaks are depicted next to them.

## 2. Fabrication of suspended atomically thin crystals

The crystals were cleaved using scotch tape to ensure a clean and flat starting surface. From this point, the materials were thinned down by successive mechanical exfoliation steps using a PDMS-based gel (Gel-Pak® PF film). The gel films were firstly inspected under the optical microscope for locating atomically thin flakes based on the optical contrast. Next, the flakes were transferred onto $SiO_2$/Si substrates with pre-fabricated micro-wells of 600nm in diameter, following a pressure-free procedure. That is, the gel film with the atomically thin flakes was placed on the substrate and, after fully engaged, it was slowly lifted up using a motorized arm. The procedure yielded atomically thin $CrI_3$ and $CrCl_3$ samples with a range of thicknesses down to the 2L. None of the 1L crystals found was suitable for the indentation experiments mainly due to their small surface area. The exfoliated samples were then transferred from the optical system to the AFM for topographic inspection. Note that both the sample preparation and AFM inspection were performed in the same chamber with an argon atmosphere to prevent the atmospheric degradation of the air-sensitive samples.

3. **Atomic force microscopy**

Topography imaging in tapping mode and nanoindentation in contact mode were conducted on the atomically thin CrX$_3$ using a Flex AFM from Nanosurf. These images were used to determine the thickness of the materials and center of the suspended regions of the CrX$_3$ for nanoindentation. Several silicon cantilevers (Tap300DLC) with a diamond-like-carbon coating on the tips were used to avoid deformation of the tip apexes under relatively high load. The spring constants of the cantilevers were determined using the thermal noise method. The tip radii were measured by scanning electron microscopy (Gemini SEM, see Fig. S2). The indentations were performed on relatively large and clean flakes for more accurate results. The load-displacement curves with obvious hysteresis were excluded from further analysis. For accuracy purposes, particular care was taken in the selection of suitable suspended crystals free of contaminations and surface cracks by both optical observation and AFM scanning. The thermal drift of the samples was taken into consideration prior to commencement of the indentation process.

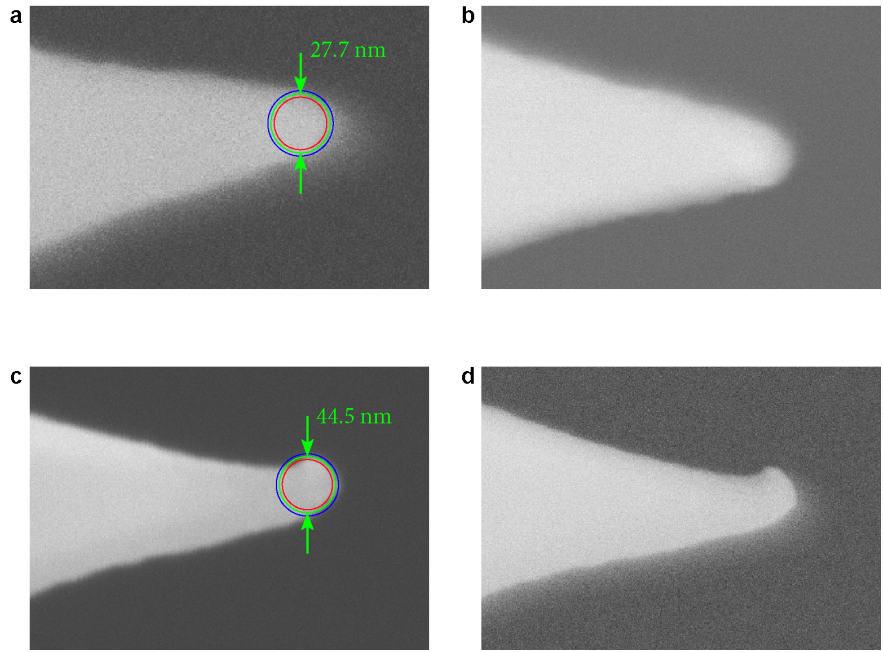

**Figure S2. Two tips are shown before and after indentation.** The first tip in panels **a,** before and **b,** after, and the second tip in **c,** before and **d,** after. The AFM indenter tip diameters were measured using a HRFESEM Zeiss GeminiSEM500. At the same time, the tips were measured before and after indenting, in order to ensure that the tip shape was not mechanically damaged or deformed.

4. **Drum resonator model**

The data extracted from the AFM indentations was fitted to a classical model for drum resonators which describes the load-displacement relation as:

$$F = \sigma_0^{2D}(\pi a)\left(\frac{\delta}{a}\right) + E^{2D}(q^3 a)\left(\frac{\delta}{a}\right)^3 \qquad (2)$$

where $E^{2D}$ and $\sigma_0^{2D}$ are the effective Young's modulus and pre-tension of the $CrX_3$ flakes with the radius ($a$) of the suspended part, stretched with vertical displacement ($\delta$) under the applied load (F); $q=1/(1.049-0.15v-0.16v^2)$ is a dimensionless coefficient related to the Poisson's ratio ($v$), which are 0.253 and 0.297 for $CrI_3$ and $CrCl_3$, respectively[33]. This model describes the indentation behavior in classical drum resonators with low bending rigidity and small pre-tension. That is, a linear trend is considered for the membrane under small loads, while a cubic trend related to the stiffness of the membrane dominates when the load increases. The conventional bulk (*i.e.*, volumetric) Young's modulus $E$ was calculated by dividing the 2D value, *i.e.* $E^{2D}$ by the sample effective thickness, the used effective thicknesses of the $CrI_3$ and $CrCl_3$ were 0.662 nm and 0.611 nm, respectively[34,35]. By using this model, we successfully extracted Young's modulus of $CrX_3$ for a range of thicknesses on both materials.

## 5. Finite elements analysis

The analysis was performed using the commercial nonlinear finite element code ABAQUS. The $CrCl_3$ and $CrCl_3$ nanosheets were modeled as axisymmetric shells with a radius of 300 nm and the initial thicknesses are 0.6114*N nm and 0.6623*N nm, respectively, where N is the number of layers. The nanoindenters were modeled as rigid spheres with radii corresponding to those used in the experiment. The model employed two-node linear axisymmetric shell elements (SAX1), with mesh densities linear variation from 0.1nm (center) to 5.0nm (outermost) along the 300nm shell radius. The tip-top layer interaction was modeled by a frictionless contact algorithm. The displacement-controlled loading with a prescribed 0.1nm per load step was applied to the spherical indenter. The nonlinear elastic constitutive behavior of both $CrCl_3$ and $CrI_3$ were assumed and it can be expressed under a uniaxial load as $\sigma=E\varepsilon+D\varepsilon^2$, where $\sigma$ and $\varepsilon$ are he the symmetric second Piola-Kirchhoff stress and the uniaxial Lagrangian strain, respectively. The Young's moduli (E) and the third-order elastic constant (D) values of $CrCl_3$ and $CrI_3$ were set to the values obtained from experimental results by using an analytical method. The nonlinear elastic behavior was implemented in ABAQUS using a previously described equivalent elastic-plastic material model[1]. The simulation loading steps corresponding to the point of membrane fracture were identified based on the fracture loads from the experiment. Therefore, fracture strength was derived as a volume average of the stress values of the elements that were directly underneath the nanoindenter at the loading step corresponding to the fracture point in the load-displacement curves obtained from the finite element methods. Full strain distributions along the membrane radius for both 2L $CrCl_3$ and $CrI_3$ are shown in Figure S3.

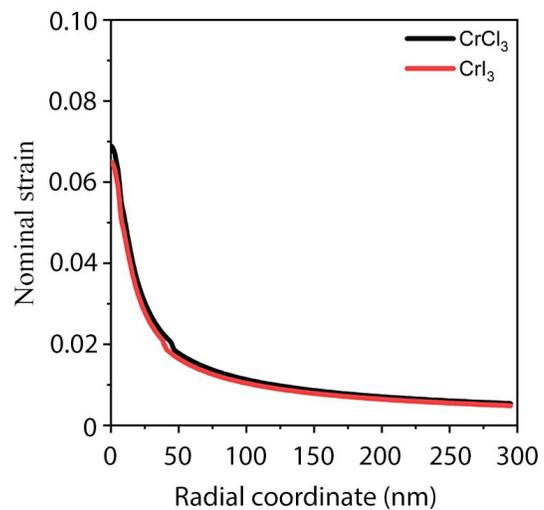

**Figure S3. Strain distribution along the radius of the 2L CrCl$_3$ and CrI$_3$ nanosheets under the fracture load**. These results were obtained from finite element simulations.

6. *Ab initio* simulations

Calculations were performed using Density Functional Theory as implemented in the VASP package[33]. Within the generalized gradient approximation (GGA), the Perdew-Burke-Ernzerhof (PBE)[34] exchange-correlation functional was used. For bilayer and bulk, Van der Waals interactions were taken into account through the many-body dispersion energy method[36]. A Hubbard correction was included through the Dudarev formulation[37]. The parameter *J* was fixed to 0.9 eV while the *U* energy was adjusted to obtain a U effective of 2.63 and 2.65 eV for $CrCl_3$ and $CrI_3$, respectively. The later U values were taken from the literature[37]. The electronic convergence criterion was set to $1 \times 10^{-7}$ eV, while the structural optimizations were performed until all forces were lower than 0.001 eV/Å. All calculations were spin polarised. Energy cut-off of 800 eV, which is considerably higher than the maximum ENMAX in the pseudopotentials (288 eV) was utilized to ensure tight convergence. For orthorhombic (LT stacking) systems with no periodicity along the z direction (monolayer, bilayer and trilayer), the k-space was sampled *via* a 6x6x1 automatically generated Gamma-centred grid. Convergence with respect to k-point sampling was addressed at this point by comparison with the mechanical properties obtained with a 9x9x1 k-point grid. As the differences in the Young's moduli were negligible (~0.1%), the 6x6x1 sampling was used throughout the entire simulation set. The k-point grid was adjusted to keep the sampling density constant whenever the dimensions of the unit cell changed, i.e., 6x4x1 for monoclinic (HT stacking) bilayers and trilayers.

We used two stackings in the simulations, namely rhombohedral (space group R$\bar{3}$) as observed at low temperature (LT) and monoclinic (space group C2/m) at high temperature (HT)[34]. Negligible variations were observed when the rhombohedral (space group R$\bar{3}$) was utilized[34,35] (see Table S1). In terms of magnetic ordering, all monolayers were considered to be FM. The bilayer and trilayer were assumed intra-layer FM and inter-layer AFM, for both materials.

In order to calculate the mechanical properties, the following settings were used:

- IBRION = 6 ; allows for the calculation of the Hessian matrix
- ISIF = 3 ; stress tensor calculated – all degrees of freedom allowed to change

- NFREE = 4 ; number of displacements per direction per ion (more than this would be excessive, 2 would be ok but not optimum, 1 should be avoided). For trilayers, NFREE=2 was required due to the computational cost.

These settings produce the 'TOTAL ELASTIC MODULI' in the OUTCAR file, namely, the matrix of elastic-stiffness coefficients 'C'. This is a 6x6 matrix relating the six independent stress and strain components (1 = XX, 2 = YY, 3 = ZZ, 4 = YZ, 5 = ZX and 6 = XY). With these coefficients, one can calculate the angle dependent in plane volumetric Young's Modulus, according to the equation[38,39]:

$$E(\theta) = \frac{C_{11} C_{22} - C_{12}^2}{C_{11} \sin\sin\theta^4 + C_{22} \cos\cos\theta^4 + \left(\frac{C_{11} C_{22} - C_{12}^2}{C_{44}} - 2 C_{12}\right) \sin\sin\theta^2 \cos\cos\theta^2} \quad (3)$$

Each of these coefficients is given volumetrically by VASP, i.e., in units of kBar. In the bulk case, they can be used directly. In the low dimensional cases, they have to be rescaled in order to 'remove' the vertical vacuum space in the unit cell. As the volume is in the denominator, the rescaling would imply multiplying the coefficient by the module of the third lattice vector ($LV_3$) projected over the vertical coordinate (Z) and dividing by the interlayer distance times the number of layers. The interlayer distance (d) is taken from our bulk optimized structure and used in the monolayer, bilayer and trilayer systems.

$$C_{rescaled} = C_{from\ VASP} \times \frac{\vec{LV_3} \cdot \hat{z}}{d \cdot N_{layers}} \quad (4)$$

As per our GGA+U calculations, the bulk interlayer distances are 5.859 and 6.698 Å for $CrCl_3$ and $CrI_3$, respectively.

|  | Young modulus (GPa) | | | |
| --- | --- | --- | --- | --- |
|  | Monoclinic (space group C2/m) | | Rhombohedral (space group R$\bar{3}$) | |
| Layers | $CrCl_3$ | $CrI_3$ | $CrCl_3$ | $CrI_3$ |
| 1L | 60.68 | 40.49 | 60.68 | 40.49 |
| 2L | 59.51 | 38.35 | 60.01 | 36.57 |
| 3L | 60.03 | 37.14 | 59.88 | 38.9 |

**Table S1**. **Young's Modulus at $\theta = 0°$ (see Figure S3) versus the number of layers.** Different stacking sequences were utilized in the simulations such as rhombohedral (space group R$\bar{3}$) which was observed at low temperature (LT) and monoclinic (space group C2/m) a high temperature (HT)[34,35].

| | | | | |
| --- | --- | --- | --- | --- |
| $CrI_3$ | Distance from the center, [nm] | 0 | 8.5 | 300 |
|  | Max contact pressure, [GPa] | 0.36 | 0.00 | 0.00 |
|  | In-plane strain, [%] | 6.09 | 4.79 | 0.00 |
| $CrCl_3$ | Distance from the center, [nm] | 0 | 8.5 | 300 |
|  | Max contact pressure, [GPa] | 0.49 | 0.00 | 0.00 |
|  | In-plane strain, [%] | 6.49 | 5.35 | 0.00 |

**Table S2**. **Values of strain and out-of-plane compression obtained by FEM.** These values of nominal strain and out-of-plane compression were obtained from FEM based on the experimental data, which has been used for sliding energy calculation by vdW-corrected density functional theory method.

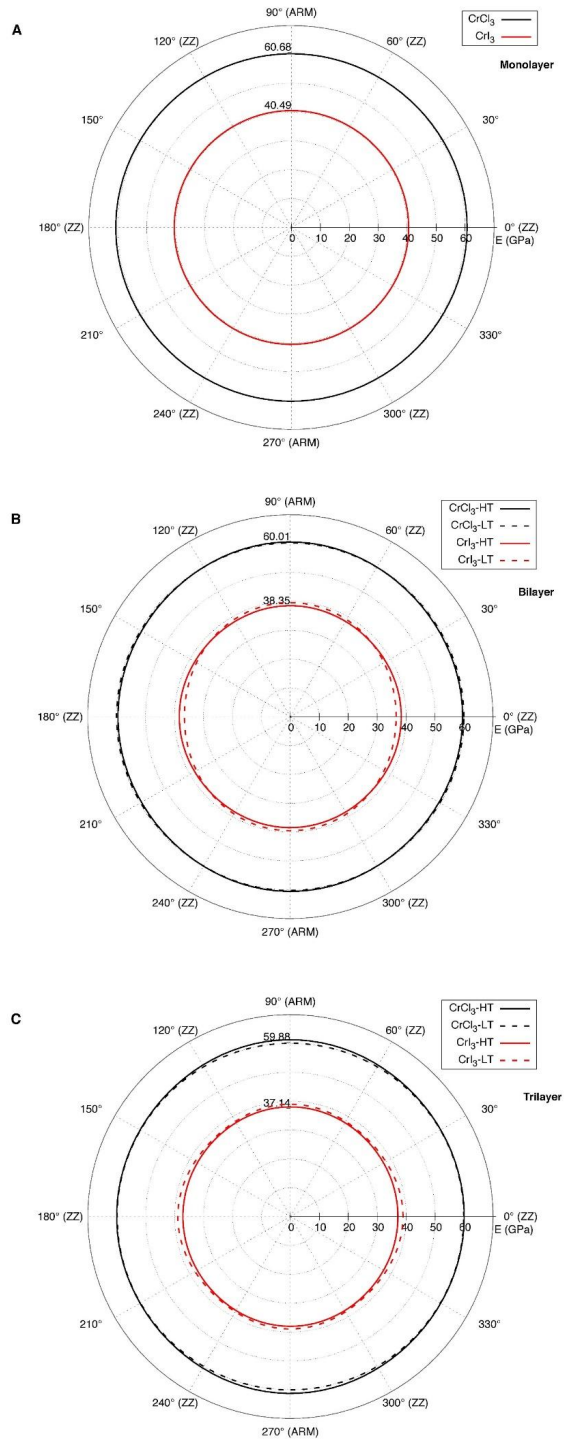

**Figure S4. Polar plots $\theta$ (º) of the Young modulus E (in GPa) of monolayer (a), bilayer (n) and trilayer (c) CrX$_3$ (X=Cl, I) using two stacking sequences: rhombohedral (space group R$\underline{3}$) at low temperature (LT) and monoclinic (space group C2/m) a high temperature (HT).**[34,35] Different orientations of the crystals along zigzag (ZZ) and armchair (ARM) are shown. Small anisotropies are observed throughout different directions.

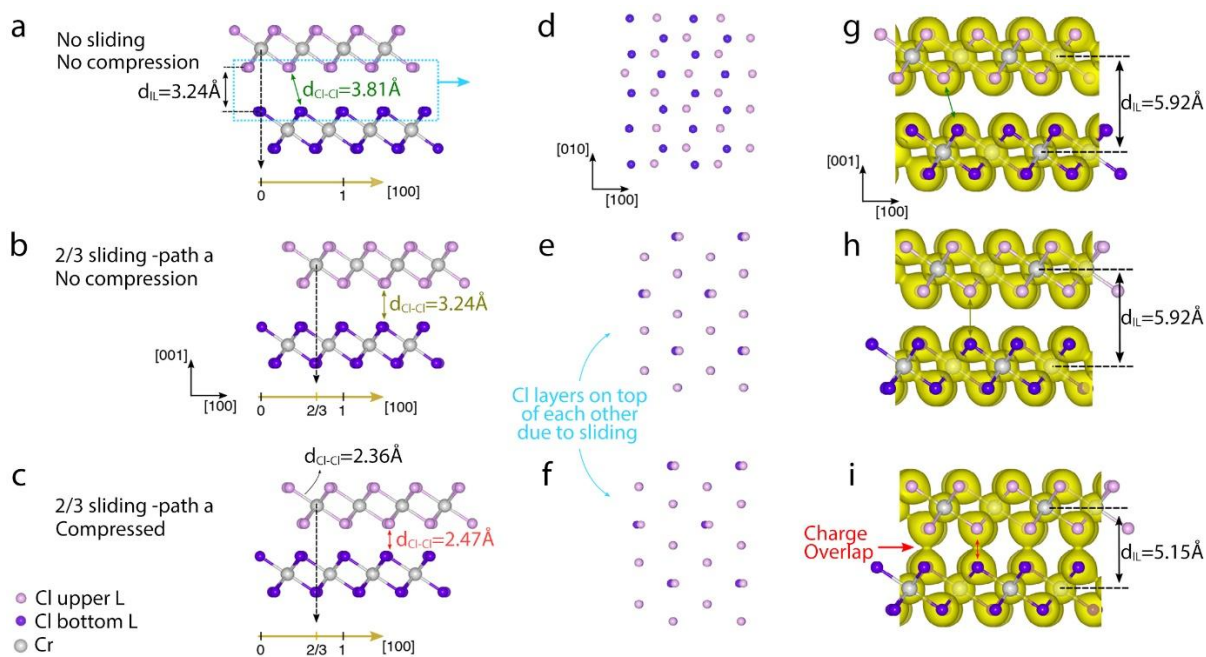

**Figure S5. Charge density calculations of CrX$_3$. a-c,** Side-, and **d-f,** top-views, respectively, of bilayer CrCl$_3$ at: **a, d,** the equilibrium position with no strain, pressure and displacement relative to the original stacking order; **b, e,** at 2/3 of the lateral shift relative to the origin at 0 along [100] with no compression; and **c, f,** at 2/3 position with applied compression. Cl atoms at the top and bottom layers are shown with distinct colors to highlight the shift. **g-i,** Charge density plots at the three situations with no compression or slide (**a, d**), with no compression but glided relative to the origin (**b, e**), and with compression (**c, f**), respectively. At 2/3 sliding along path-a ([100]) the chalcogenide atoms across the van der Waals (vdW) gap end up right on top of each other. That is, the lower Cl plane of the upper layer and the upper Cl plane of the lower layer have the same x and y coordinates. This results in the shortening of the Cl-Cl distance across the vdW gap, and justifies why the energy increases in that particular sliding condition (in general, regardless of the pressure). This doesn't happen with other stackings or glidings. When hydrostatic pressure is applied in the system, the reduction of the interlayer distance brings the Cl atoms even closer. This ended up in covalent-like distances, and a strong overlapping charge density. Similar analysis applies for CrI$_3$ not shown.

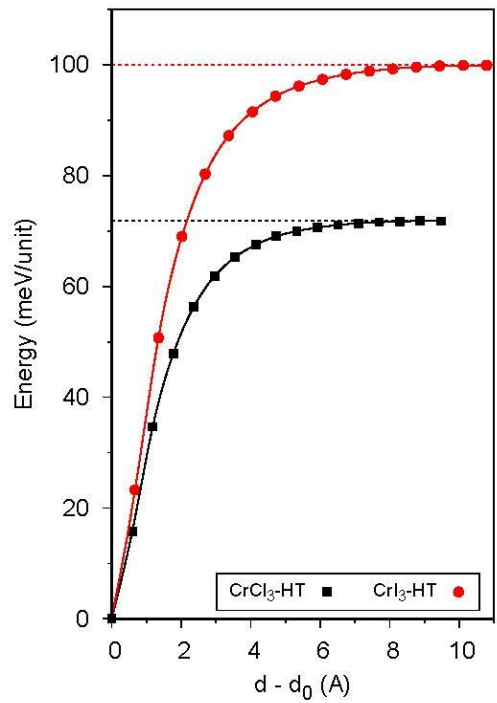

**Figure S6. Calculated energy to separate bilayer into two monolayers using DFT**. $d_0$ is the equilibrium interlayer distance and d is the separation between layers of $CrX_3$.

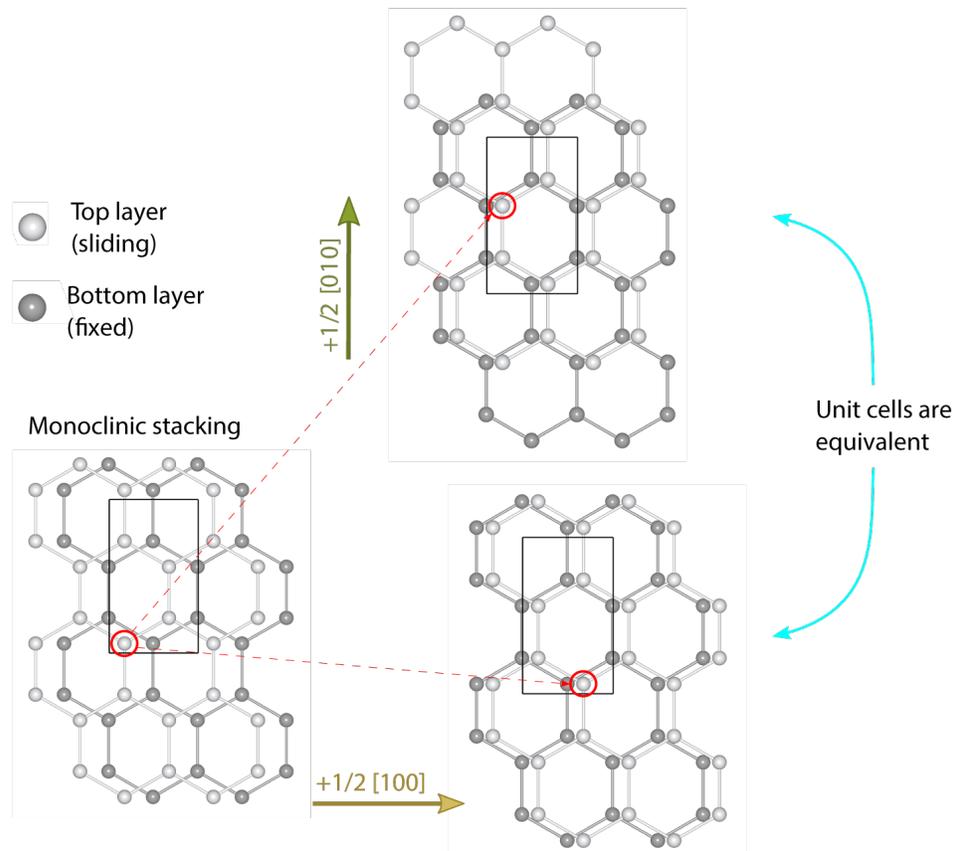

**Figure S7. Schematic of the equivalence of sliding process from 1/2 to 2/3 along [100] or [010].** Both situations resulted in similar unit cells.

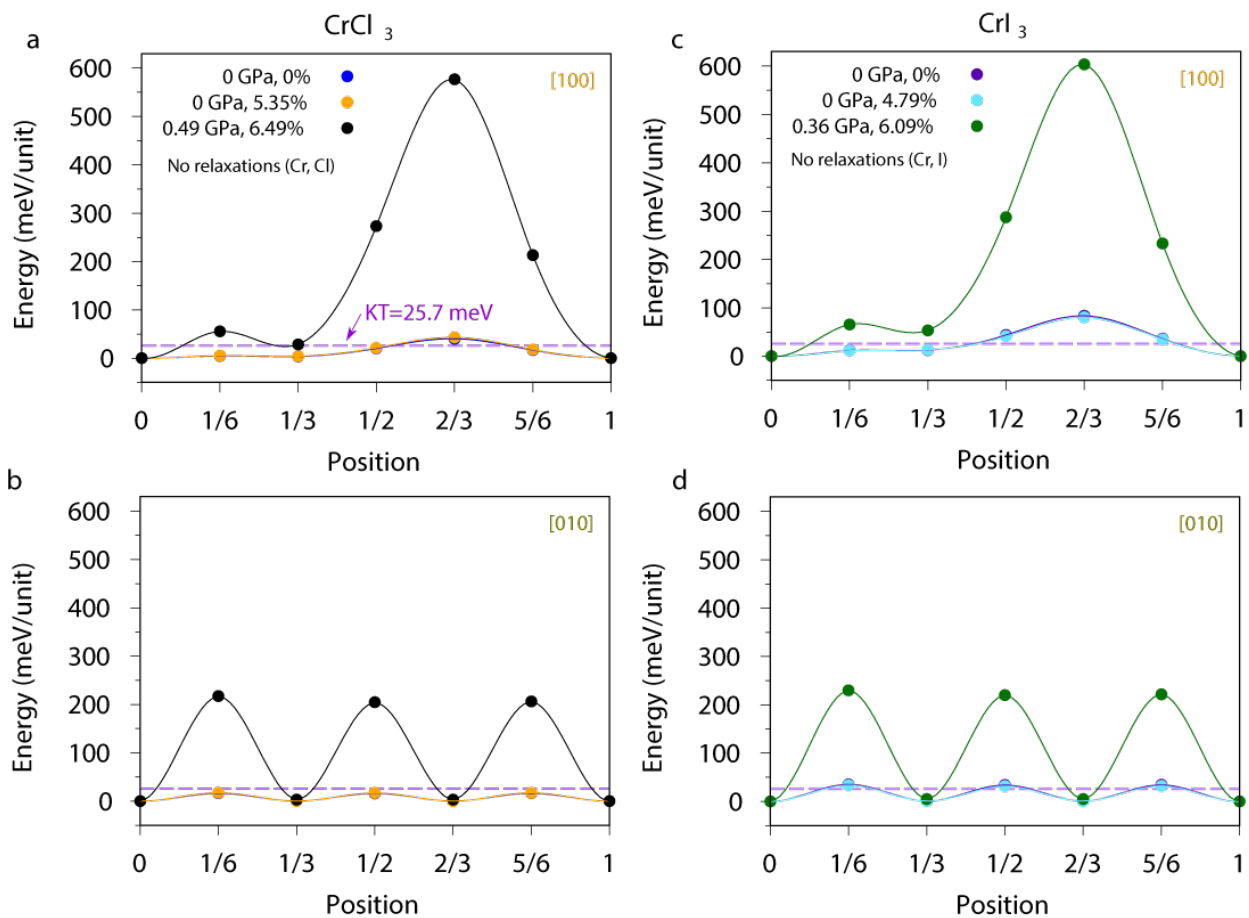

**Figure S8. Calculated energy barriers for CrCl3 and CrI3 along [100] and [010].** This is similar as Figure 3 in the main text but with no relaxation performed on the Cr, Cl, I atoms.

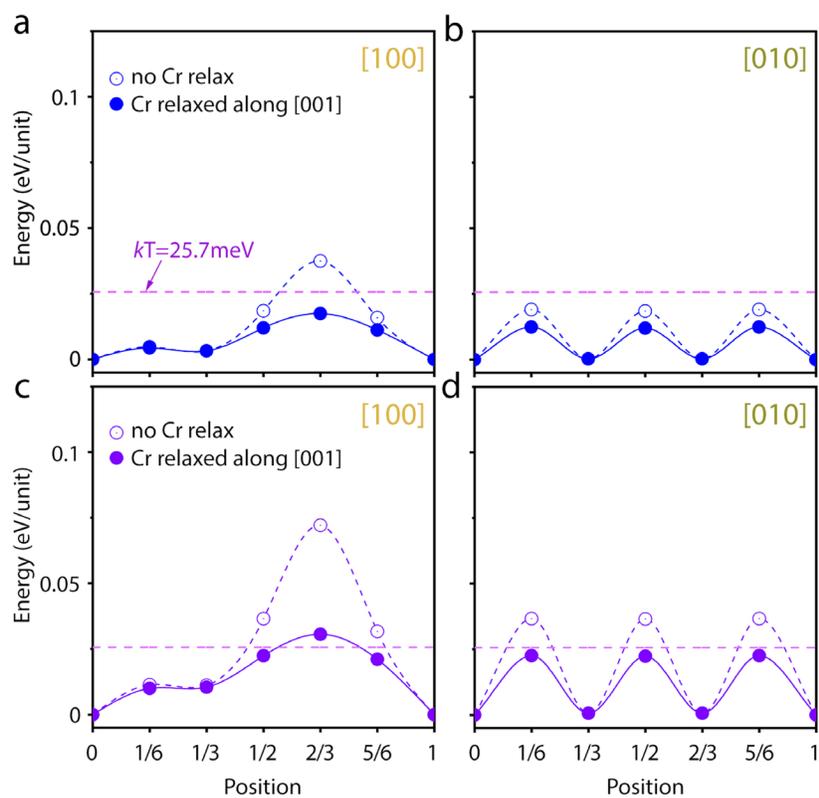

**Figure S9. Calculated energy barriers at 0 GPa and 0% strain for CrCl$_3$ (a, b) and CrI$_3$ (c,d) taking into account relaxations of the Cr atoms at both layers.** The halide atoms are previously relaxed as shown in the results Figure 3. The [001] orientation is pointed out of the plane.

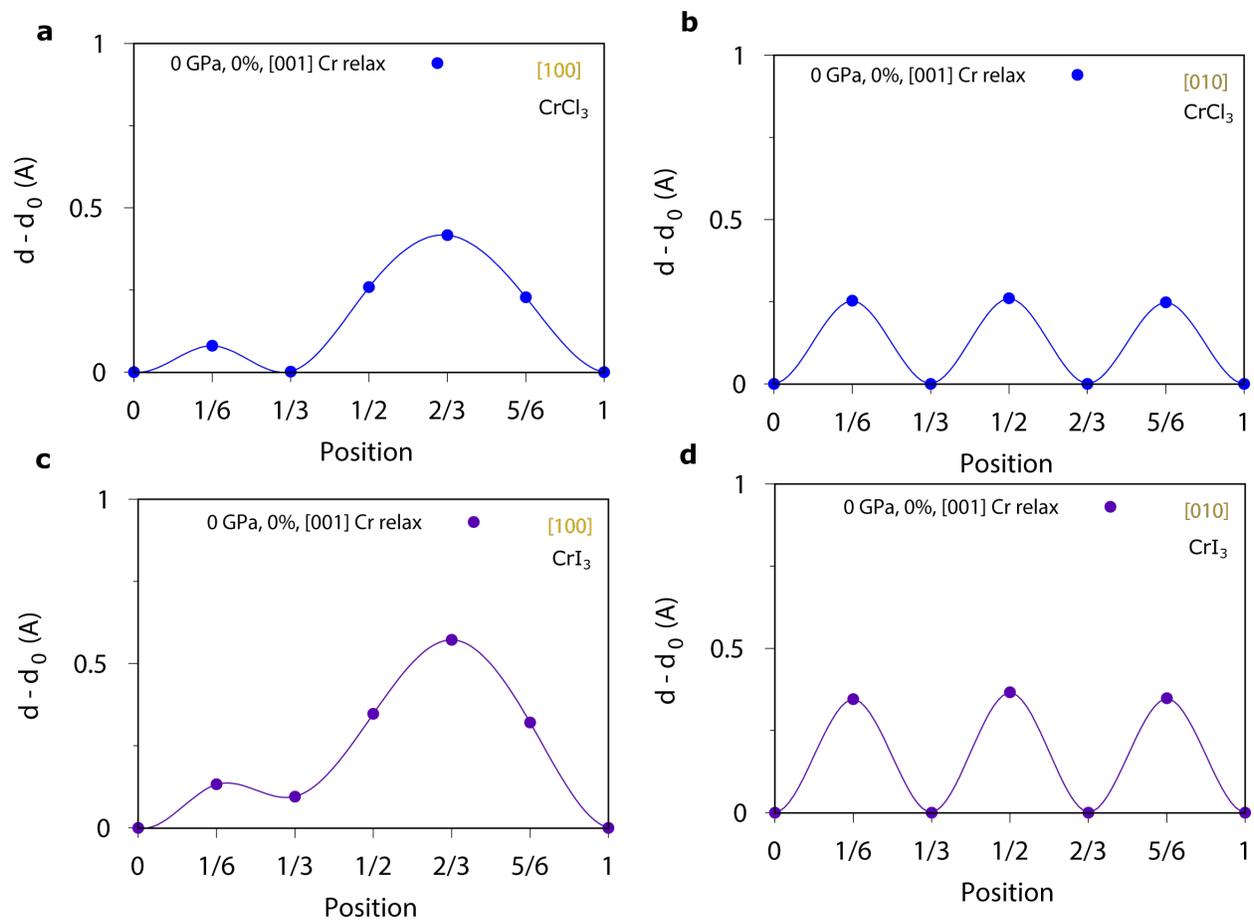

**Figure S10: Variation of the interlayer distance at the different positions of the sliding path:** along [100] and [010] for $CrCl_3$ (a, b) and $CrI_3$ (c, d) at 0 GPa and 0% strain. Cr atoms are allowed to relax along the [001] direction. $d_0$ corresponds to the equilibrium position.

## 7. Brittle nature of few-layer CrX3

As proof of the brittle nature of the free-standing CrX$_3$, Figure S11 provides evidence of reversible load-displacement curves in a free-standing membrane of few-layer CrX$_3$. In this experiment multiple consecutive indentations were performed in the same membrane avoiding reaching loads, which exceed the rupture point.

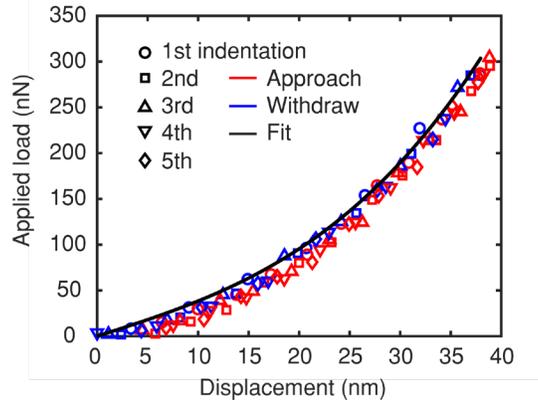

**Figure S11. Brittle nature of few-layer CrCl$_3$.** Set of five consecutive indentations without fracture on the same membrane (CrCl$_3$ ~ 10 nm). The data points have been downsampled to differentiate between the different indentations. It can be seen that the approach and withdrawal curves follow the same path and are reproducible, indicating that the behavior of the membrane is elastic and not plastic or ductile.

## 8. Enhanced plasticity in multilayer CrX$_3$

| Materials | E$_s$ (meV/f.u.) | E$_c$ (meV/f.u.)) | E$_{in}$ (GPa) | Ξ factor (GPa$^{-1}$) |
|---|---|---|---|---|
| CrCl$_3$ | 19.01500 | 71.64250 | 62.1±4.8 | 0.06104±0.00471 |
| CrI$_3$ | 36.68625 | 99.73875 | 43.4±4.4 | 0.06330±0.00642 |

**Table S3. Calculated the deformability factor for CrX$_3$ materials.**

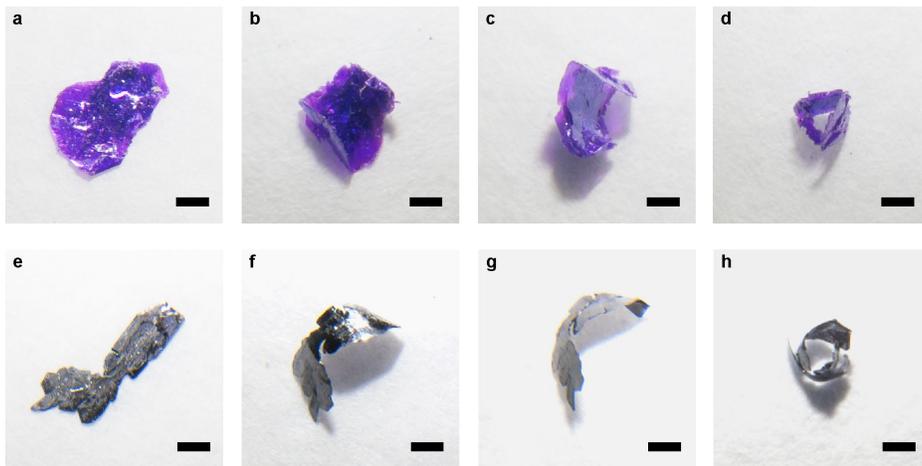

**Figure S12. Macroscopic bending test on bulk crystals. a-d,** CrCl$_3$, **e-h,** CrI$_3.$ All scale bars are 1 mm in size.

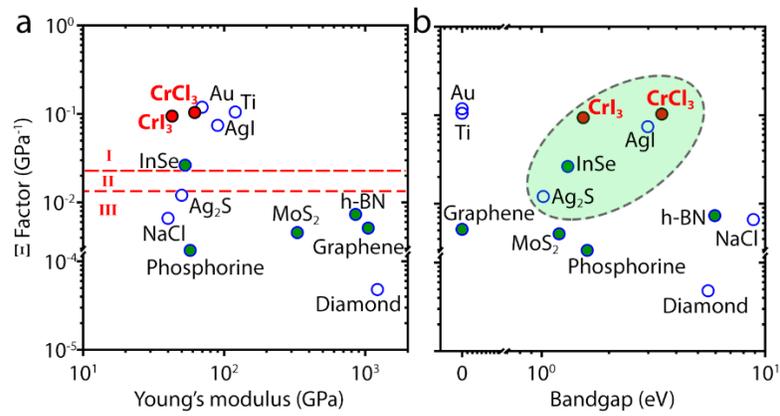

**Figure S13. Deformability factor calculated for multilayer CrX$_3$ with both Cr and halide atoms relaxed during computation.** Deformability factor dependence a, on Young Modulus and b, on the bandgap for the same materials. The deformability factor was determined with Cr atoms relaxed along [001] and halides (X=Cl, I) are relaxed in-plane. The legends to the plots are the same with Figure 5 in the main text.